\documentclass[letter,traditabstract]{aa}

\usepackage{graphicx}
\usepackage{amssymb}
\usepackage{amsmath}
\usepackage{url}
\usepackage{natbib}
\bibpunct{(}{)}{;}{a}{}{,}

\newcommand{\onefig}[3]{%
  \begin{figure}%
    \centerline{\resizebox{\hsize}{!}{\includegraphics*{#1}}}%
    \caption{#3}\label{#2}%
  \end{figure}%
}
\newcommand{\sect}[1]{Sect.~\ref{#1}}

\newcommand{\fig}[1]{Fig.~\ref{#1}}

\newcommand{\eq}[1]{Eq.~(\ref{#1})}

\newcommand\lmax{\ell_{\mathrm{max}}}

\newcommand\lm{\ell m}
\newcommand\alm{$a_{\ell m}$}
\newcommand\order[1]{${{\cal O}\! \left( #1 \right)}$}
\newcommand\beq{\begin{equation}}
\newcommand\eeq{\end{equation}}

\begin{document}

\title{Fast calculation of the Fisher matrix for cosmic microwave
  background experiments}

\titlerunning{Fast calculation of the Fisher matrix}

\author{Franz Elsner\inst{1}
  \and
  Benjamin D. Wandelt\inst{1,2}}

\institute{Institut d'Astrophysique de Paris, UMR 7095, CNRS -
  Universit\'e Pierre et Marie Curie (Univ Paris 06), 98 bis blvd
  Arago, 75014 Paris, France\\
  \email{elsner@iap.fr}
  \and
  Departments of Physics and Astronomy, University of Illinois at
  Urbana-Champaign, Urbana, IL 61801, USA}

\date{Received \dots / Accepted \dots}

\abstract{The Fisher information matrix of the cosmic microwave
  background (CMB) radiation power spectrum coefficients is a
  fundamental quantity that specifies the information content of a CMB
  experiment. In the most general case, its exact calculation scales
  with the third power of the number of data points $N$ and is
  therefore computationally prohibitive for state-of-the-art
  surveys. Applicable to a very large class of CMB experiments without
  special symmetries, we show how to compute the Fisher matrix in only
  \order{N^2\log N} operations as long as the inverse noise covariance
  matrix can be applied to a data vector in time \order{\lmax^3\log
    \lmax}. This assumption is true to a good approximation for all
  CMB data sets taken so far. The method takes into account common
  systematics such as arbitrary sky coverage and realistic noise
  correlations. As a consequence, optimal quadratic power spectrum
  estimation also becomes feasible in \order{N^2\log N} operations for
  this large group of experiments. We discuss the relevance of our
  findings to other areas of cosmology where optimal power spectrum
  estimation plays a role.}


\maketitle

\section{Introduction}
\label{sec:intro}

Observations of the cosmic microwave background (CMB) radiation have
proven to be a cornerstone of modern cosmology, leading to a vigorous
experimental effort to measure its anisotropies
\citep[e.g.,][]{1992ApJ...396L...1S, 2002ApJ...571..604N,
  2003ApJS..148....1B, 2004ApJ...600...32K, 2011ApJ...739...52D,
  2011ApJ...743...28K}. Whilst the limited angular resolution of the
first CMB experiments naturally restricted the data size, observations
obtained with Planck, the third generation CMB satellite experiment,
and high-resolution ground-based experiments will soon deliver maps
with as many as \order{10^7} pixels \citep{2011A&A...536A...1P},
challenging the performance of data analysis tools.

The information on the most fundamental cosmological parameters given
by a CMB sky map is fully contained in the much smaller set of angular
power spectrum coefficients \citep{1996PhRvD..54.1332J,
  1997MNRAS.291L..33B}. This makes the power spectrum a convenient
intermediate stage product in the analysis chain. For the lossless
extraction of the parameters in a subsequent step, however, a thorough
characterization of the statistical properties of the power spectrum
coefficients is necessary.

As one of the most important objects in statistics, the Fisher
information matrix quantifies the ability to constrain a set of
parameters by means of an experiment. In addition, it reflects the
(possibly complicated) correlation structure among them. Though
formally only applicable to (multivariate) Gaussian variables, the
Fisher information matrix has been put to use in a wide variety of
cosmological contexts, e.g., galaxy surveys
\citep[e.g.,][]{2000MNRAS.312..285H, 2003ApJ...598..720S},
gravitational wave astronomy \citep[e.g.,][]{2005PhRvD..71h4025B,
  2008PhRvD..77d2001V}, weak lensing surveys
\citep[e.g.,][]{1999ApJ...514L..65H, 2008MNRAS.389..173K}, and to
optimize numerical quadrature \citep[e.g.,][]{2011MNRAS.417....2S},
etc.

Unfortunately, the exact calculation of the Fisher matrix for the CMB
power spectrum of present-day experiments is numerically intractable:
in the most general case, the time complexity of suitable algorithms
is \order{N^3}, where $N$ is the number of pixels of the CMB data map
\citep{1999PhRvD..59b7302B}. To overcome this problem, approximate
methods have been proposed to speed up the calculation. These methods
may rely on the use of Monte Carlo averages to estimate expectation
values \citep{1997PhRvD..55.5895T}, or numerical differentiation of
the likelihood function \citep{2006JCAP...10..013P}. However, they are
plagued by stability issues and unable to calculate off-diagonal
elements to reasonable precision. Thus, an exact scheme to evaluate
the Fisher matrix with moderate computational cost has remained
elusive until now.

The paper is organized as follows. In \sect{sec:torus}, we provide a
short overview of the mathematical framework that our approach is
based on. In \sect{sec:fisher}, we then introduce an efficient way of
calculating the Fisher information matrix. We first consider
experiments with a simplified scanning strategy to introduce the main
ideas, and then generalize our findings to more generic CMB
experiments. Finally, we summarize our results in \sect{sec:summary}.

\section{The likelihood function on the ring torus}
\label{sec:torus}

For an arbitrary CMB experiment with Gaussian noise, the likelihood
function ${\cal L}$ of the data $d$ is given by
\begin{multline}
\label{eq:likelihood}
{\cal L}(C_{\ell} \vert d) = \frac{1} {\sqrt{ \vert 2 \pi
    (\mathbf{S}(C_\ell)+\mathbf{N}) \vert}} \\
\times \exp \left[ {-\frac{1}{2} \, d^{\dagger} (\mathbf{S}(C_\ell) +
    \mathbf{N})^{-1} d} \right] \, ,
\end{multline}
where we have introduced the noise covariance matrix, $\mathbf{N}$,
and the signal covariance matrix, $\mathbf{S}$, a function of the CMB
power spectrum coefficients $C_{\ell}$.

For an efficient yet still exact evaluation of the likelihood, we
start out by considering a CMB experiment that scans the sky on
iso-latitude circles. The foundations of this method are explained in
detail in \citet{2003PhRvD..67b3001W}. Imposing this restriction on
the scanning strategy allows us to map the resulting time-ordered data
(TOD) structure onto the ring torus. Owing to the periodic boundary
conditions, this method enables us to work in the Fourier basis, where
the algebraic expressions are simpler.

Quantifying the signal covariance structure in Fourier space, we first
write the noise-free sky temperature in terms of spherical harmonic
coefficients of the signal map, \alm,
\beq
 \label{eq:signal_torus}
 T^{S}_{r p} = \sum_{\ell} a_{\ell r} \, d^{\ell}_{r
   p}(\theta_{\mathrm{s}}) \, X_{\ell p} \, ,
\eeq
where the index $p$ runs over the Fourier modes in the direction of
the rings, and $r$ specifies the index for the cross-ring direction.
In \eq{eq:signal_torus}, we apply the definition of the Wigner
rotation matrix to introduce the real quantity $d$
\beq
D^{\ell}_{m m'}(\phi_2,\theta,\phi_1) = \mathrm{e}^{-\mathrm{i} m \phi_2}
\, d^{\ell}_{m m'}(\theta) \, \mathrm{e}^{-\mathrm{i} m' \phi_1} \, ,
\eeq
where we choose $\theta = \theta_{\mathrm{s}}$, the constant latitude
of the experiment's spin axis. We also make use of the rotated beam
$X$ according to
\beq
X_{\ell m} = \sqrt{\frac{2 \ell + 1}{4 \pi}} \sum_{m'} d^\ell_{m
  m'}(\theta_{\mathrm{o}}) \, b_{\ell m'}^{\ast} \, ,
\eeq
where the $b_{\lm}$ are the expansion coefficients of the beam
function in spherical harmonics, and the Wigner small $d$ matrix is
evaluated at the opening angle of the scanning circles,
$\theta_{\mathrm{o}}$. We note that this framework allows an exact
treatment of arbitrarily shaped beam patterns. The signal covariance
matrix $\langle T^{S}_{r p} T^{S \, \ast}_{r' p'} \rangle$ is then
given by
\beq
\mathbf{S}_{r p r' p'} = \delta_{r r'} N^2 \sum_{\ell} C_{\ell} \,
d^{\ell}_{r p} X_{\ell p} \, d^{\ell}_{r' p'} X_{\ell p'}^{\ast} \, ,
\eeq
where the Kronecker delta ensures the block diagonal structure of the
signal covariance matrix in Fourier space.

After having specified the signal properties, we now quantify the
noise correlations. Assuming a stationary Gaussian process, we can
describe the noise properties in the TOD domain by a power spectrum
$P(k)$. We stress that this ansatz enables the exact treatment of
correlated noise, a common systematic effect in real-world
experiments. The noise covariance matrix $\langle T^{N}_{r p} T^{N \,
  \ast}_{r' p'} \rangle$ simplifies to
\begin{multline}
\mathbf{N}_{r p r' p'} = \delta_{r r'} \frac{1}{N_{\mathrm{r}}
  N_{\mathrm{p}}^2} \sum_{m, m' = 0}^{N_{\mathrm{p}} - 1}
\mathrm{e}^{-\frac{2 \pi \mathrm{i}} {N_{\mathrm{p}}} (p m - p' m')} \\
\times \sum_{\Delta = 0}^{N_{\mathrm{r}} - 1} \mathrm{e}^{-\frac{2 \pi
    \mathrm{i}}{N_{\mathrm{r}}} \Delta r} C(\Delta, m, m') \, ,
\end{multline}
where $N_{\mathrm{r}}$ is the number of rings in the data set, and
$N_{\mathrm{p}}$ the number of pixels per ring. We have introduced an
auxiliary function $C(\Delta, m, m')$, which is defined as
\beq
C(\Delta, m, m') = \sum_{k=0}^{N-1} P(k) \, e^{-\frac{2\pi \mathrm{i}}
  {N} k (N_r \Delta + m - m')} \, .
\eeq
In analogy to $\mathbf{S}$, we also find the noise covariance matrix
$\mathbf{N}$ to be block diagonal in Fourier space.

\section{Calculating the Fisher matrix on the ring torus}
\label{sec:fisher}

We now propose a strategy for the exact calculation of the Fisher
matrix in only \order{N^2\log N} operations. Given the definition of
the Fisher matrix as the covariance of the score function, from
\eq{eq:likelihood}, we obtain \citep{1997PhRvD..55.5895T}
\begin{align}
 \label{eq:fisher}
 F_{\ell_{1} \ell_{2}} &= - \left \langle \frac{\partial^2 \ln {\cal
     L}}{\partial C_{\ell_{1}} \partial C_{\ell_{2}}} \right \rangle
 \nonumber \\
 &= \frac{1}{2} \, \mathrm{tr} \left[ \mathbf{C}^{-1}
   \mathbf{P}^{\ell_{1}} \mathbf{C}^{-1} \mathbf{P}^{\ell_{2}} \right] \, ,
\end{align}
where $\mathbf{C} = \mathbf{S} + \mathbf{N}$, and $\mathbf{P}^{\ell} =
\partial \mathbf{C} / \partial C_{\ell}$. The evaluation of
\eq{eq:fisher} in its general form written here takes \order{N^3}
operations \citep{1999PhRvD..59b7302B}.

\subsection{Idealized ring-torus scanning strategy}

We first consider an idealized experiment as described in the last
section. For the calculation of the Fisher matrix, we need to compute
the inverse of the covariance matrix, $\mathbf{C}^{-1}$, which can be
done brute force in \order{N^2} operations owing to its block diagonal
structure. The second ingredient, according to \eq{eq:fisher}, is the
derivative of the covariance matrix with respect to the power spectrum
coefficients. On the ring torus, we can calculate the derivative for
each block $r$ separately as
\begin{align}
\label{eq:cov_deriv}
\left( \frac{\partial \mathbf{C}_{r}}{\partial C_{\ell}} \right)_{p \,
  p'}&= \left( \frac{\partial \mathbf{S}_{r}}{\partial C_{\ell}}
\right)_{p \, p'} \nonumber \\
&= N^2 \, d^{\ell}_{r p} X_{\ell p} \, d^{\ell}_{r p'} X_{\ell p'}^{\ast}
\nonumber \\
&= q^{\ell}_{r p} \, q^{\ell \, \ast}_{r p'} \, ,
\end{align}
that is, each block of the matrix is a rank one object and can
therefore be written as the outer product of a single vector
$q$. Introducing the decomposition $\mathbf{P}^{\ell}_{r} =
q^{\ell}_{r} \, q^{\ell \, \dagger}_{r}$, \eq{eq:fisher} simplifies
considerably to
\begin{align}
\label{eq:fisher_torus}
F_{\ell_{1} \ell_{2}} &= \frac{1}{2} \sum_{r} \mathrm{tr} \left[
  \mathbf{C}^{-1}_{r} q^{\ell_{1}}_{r} q^{\ell_{1} \, \dagger}_{r}
  \mathbf{C}^{-1}_{r} q^{\ell_{2}}_{r} q^{\ell_{2} \, \dagger}_{r}
  \right] \nonumber \\
&= \frac{1}{2} \sum_{r} \vert q^{\ell_{1} \, \dagger}_{r}
\mathbf{C}^{-1}_{r} q^{\ell_{2}}_{r} \vert^2 \, .
\end{align}
This exact rank-1 representation of the derivative of the signal
covariance matrix in the ring-torus Fourier domain is the key
ingredient to our fast algorithm. We note that the analogous
derivative in the spherical harmonic domain is also block diagonal but
with blocks of rank $2\ell+1$ \citep{1997PhRvD..55.5895T}. We now show
how to exploit this low-rank representation.

For each of the \order{\lmax} independent blocks of the covariance
matrix, we perform the following calculations. We first evaluate the
$\lmax$ matrix-vector products $\mathbf{C}^{-1} q$, at a numerical
cost of \order{\lmax^2}. We then accumulate each of the $\lmax^2$
entries of the Fisher matrix by a simple vector dot product, with a
\order{\lmax} scaling. The overall time complexity of the algorithm
therefore amounts to ${\cal O}\! \left( \lmax^4 \right) \approx {\cal
  O}\! \left( N^2 \right)$ operations.

To confirm the scaling behavior as claimed above, we did a full
numerical implementation of the algorithm. In \fig{fig:scaling}, we
show the timings we obtained for different values of $\lmax$. The
measured runtime clearly follows the predicted time complexity
\order{\lmax^4}. The tests were carried out on a cluster of Intel
Xenon processors with a 3~GHz clock rate using 256 CPU cores.

\onefig{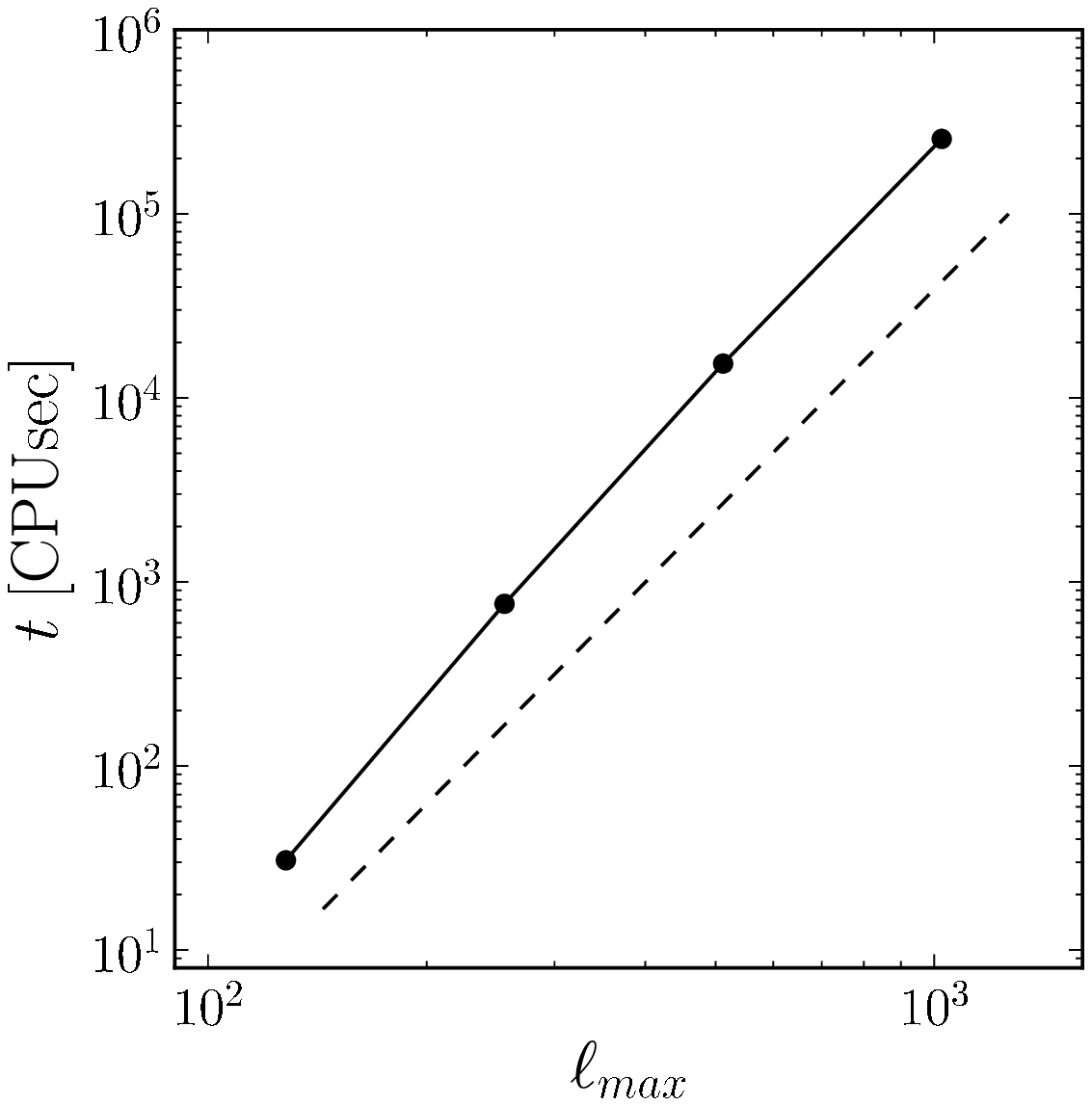}{fig:scaling}{Fast calculation of the Fisher
  matrix. For several values of $\lmax$, we show the measured runtime
  of our algorithm (\emph{filled circles}). To guide the eye, we
  overplot a $\lmax^4$ power-law (\emph{dashed line}).}

\subsection{Generalization beyond ring-torus scans and arbitrary sky
  coverage}

After having outlined the algorithm for the specific case of a
simplified scanning strategy, we now discuss how to generalize the
results to a generic CMB experiment. To this end, we propose a
resampling of the input data to a grid that can be mapped onto the
ring torus (e.g., an ECP grid, where $\theta_{\mathrm{s}} =
\theta_{\mathrm{o}} = \pi/2$). This has the advantage that the
simplifying signal covariance properties, as described above, are
still fulfilled, i.e.\ \eq{eq:cov_deriv} still holds. Unfortunately,
as the grid no longer follows the scanning strategy, the possibility
of achieving an exact treatment of asymmetric beams at no additional
computational costs is lost. In addition, the noise properties may be
more complicated, owing to, e.g., anisotropically distributed
observation time, or the presence of a Galactic mask. As a result, an
explicit expression for the covariance matrix $\mathbf{C}$ may not
exist. Luckily, we only have to evaluate products of inverse
covariance matrix and sky map. This problem is standard in CMB data
analysis, and iterative solvers have been successfully developed for
this problem \citep{2007PhRvD..76d3510S}.

As the block diagonal structure of $\mathbf{C}$ is destroyed, we solve
\beq
F_{\ell_{1} \ell_{2}} = \frac{1}{2} \vert q^{\dagger}_{\ell_{1}}
\mathbf{C}^{-1} q_{\ell_{2}} \vert^2 \, ,
\eeq
by first finding solutions to the $\lmax$ equations $x =
\mathbf{C}^{-1} q$ iteratively, and then contracting the outcome with
a simple dot product. Once the calculation has been completed
successfully, the resulting Fisher matrix is \emph{independent} of the
pixelization used for the actual computation.

The overall \order{N^2} time complexity of the algorithm can be
sustained, if the solver scales as \order{\lmax^3}. In the most
general case, where the full characterization of the noise properties
requires a dense $N \times N$ noise covariance matrix, the numerical
costs increase to ${\cal O}\! \left( \lmax^5 \right) \approx {\cal
  O}\! \left( N^{5/2} \right)$ operations. Even worse, the memory
requirement to store the matrix rises to \order{N^2}, reaching
thousands of terabytes for Planck-sized data.

Luckily, all carefully designed CMB experiments allow either a
structured or a sparse representation of the noise covariance
matrix. In the presence of an anisotropic yet uncorrelated noise and
arbitrary sky coverage, for example, it reduces to a diagonal matrix
in a real space representation. This ansatz made an efficient inverse
variance weighting of WMAP data possible
\citep{2011ApJS..192...18K}. Under very general circumstances, the
noise covariance can be written in a mixed real-space and time-ordered
domain representation, e.g.\
$\mathbf{N}^{-1}=\mathbf{A}^{T}\mathbf{N}_{\mathrm{TOD}}^{-1}\mathbf{A}$,
where $\mathbf{A}$ is the pointing matrix and
$\mathbf{N}_{\mathrm{TOD}}$ is the noise covariance in the
time-ordered domain. For essentially all past and current CMB
experiments, the noise is piecewise stationary in the time-ordered
domain, which means that $\mathbf{N}_{\mathrm{TOD}}$ is block-Toeplitz,
and the matrix-vector product of $\mathbf{N}^{-1}$ with the data takes
${\cal O}(N_{\mathrm{TOD}}\log N_{\mathrm{TOD}})$ operations. Since
there are at most ${\cal O}(\lmax^{3})$ orientations in which a CMB
experiment can scan the sky, $N_{\mathrm{TOD}} \sim \lmax^{3}$. As a
consequence, the calculation of the Fisher matrix in \order{N^2\log N}
operations becomes feasible.

A straightforward treatment of asymmetric beams is only possible if
the data grid follows the scanning strategy directly. Although this is
not the case in the general setup discussed above, we note that
asymmetric beams can be included into the analysis at additional
computational costs of \order{m^2_{\mathrm{max}}}, determined by the
azimuthal structure of the beams. For example, the treatment of
elliptical Gaussian beams increases the numerical complexity by a
factor of four.

\section{Discussion and conclusions}
\label{sec:summary}

The Fisher information matrix is a fundamental quantity in statistics,
reflecting the predictive power of the experiment under
consideration. Despite its importance, thorough systematic
Fisher-matrix-based studies of experiments, designed to measure the
CMB power spectrum, have not yet been conducted. The reason for this
shortcoming can be found in the associated computational expenses:
general algorithms show a time complexity of \order{N^3}, where $N$ is
the number of pixels of the survey, rendering an analysis for state of
the art experiments impossible \citep{1999PhRvD..59b7302B}.

Here, we have presented a new method for the exact calculation of the
Fisher matrix of the CMB power spectrum coefficients in only
\order{N^2\log N} operations. To do so, we first considered
experiments with a specific scanning strategy, where the sky is
observed on iso-latitude circles. This restriction has allowed us to
map the TOD onto the ring torus in order to take advantage of the
symmetries of this manifold. We then cast the equation for the CMB
likelihood function into Fourier space and found that both the signal
and the noise covariance matrices are block diagonal
\citep{2003PhRvD..67b3001W}. For each block, the derivative of the
signal covariance matrix with respect to a single power spectrum
coefficient was also found to be a rank one matrix. As a result, we
made significant progress in ensuring that the equations to calculate
the Fisher matrix exactly simplify, requiring only \order{N^2}
operations to evaluate.

We then relaxed the assumption of a simplified scanning strategy. If
iterative methods allowed us to calculate inverse-variance-weighted
sky maps in \order{\lmax^3\log \lmax} operations, we showed that a
\order{N^2\log N} time complexity of our algorithm could be
realized. To this end, the data were resampled onto an ECP grid, which
is homeomorphic to the ring torus, where the properties of the signal
covariance matrix were still simplified. Our fast method is applicable
to a very general class of experiments, including, to a good
approximation, all CMB data sets taken up to now.

We consider a fast means of calculating the Fisher information matrix
for CMB experiments of great importance. This new method not only
allows us to study quantitatively the impact of the common systematics
of realistic CMB experiments, such as partial sky coverage, asymmetric
beams, or correlated noise, on the power spectrum. The Fisher matrix
also appears as a normalization factor for unbiased and optimal power
spectrum estimators \citep{1997PhRvD..55.5895T}. Since its fast
calculation has not been possible, until now, only approximative
pseudo-$C_{\ell}$ estimators have been efficient enough to be
applicable to more recent CMB experiments
\citep[e.g.,][]{2001ApJ...561L..11S, 2001PhRvD..64h3003W,
  2002ApJ...567....2H}. The new scheme outlined here has the potential
to lift this restriction.

For the presentation of our method, we explicitly considered an
experiment designed to measure the CMB power spectrum. However, the
described framework is general enough to be of value in other fields
of application, where the statistical properties of an isotropic
signal in spherical geometry are investigated. These applications may
include, but are not limited to, the analysis of galaxy angular power
spectra \citep{2005MNRAS.360L..82R, 2008ApJ...686L...1L}, or weak
lensing tomography \citep{2004ApJ...600...17B, 2004MNRAS.348..897T}.

\begin{acknowledgements}
  BDW was supported by the ANR Chaire d'Excellence and NSF grants AST
  07-08849 and AST 09-08902 during this work. FE gratefully
  acknowledges funding by the CNRS. This research was supported in
  part by the National Science Foundation through XSEDE resources
  provided by the XSEDE Science Gateways program under grant number
  TG-AST100029.
\end{acknowledgements}

\bibliographystyle{aa}
\bibliography{literature}

\end{document}